\documentclass[
    ,final            % use final for the camera ready runs
  ]
  {aipproc}

\layoutstyle{6x9}

\def\be{\begin{equation}}
\def\ee{\end{equation}}
\def\bea{\begin{eqnarray}}
\def\eea{\end{eqnarray}}

\def\nnw{\nonumber \\ [.2cm]}
\def\vsp#1{\vspace*{#1}}
\def\hsp#1{\hspace*{#1}}

\def\cH{{\cal H}}
\def\cK{{\cal K}}  
\def\cL{{\cal L}}

\def\part{\partial}
\def\tfrac#1#2{{\textstyle{\frac{#1}{#2}}}}
\def\half{\tfrac{1}{2}}

\def\mn{{\mu\nu}}
\def\mnr{{\mu\nu\rho}}

\def\sqrtg{\sqrt{|g|}}

\begin{document}

\title{The Palatini formalism for higher-curvature gravity theories}


\author{Mar Bastero-Gil}{
  address={Departamento de F\'isica Te\'orica y del Cosmos\\
           and Centro Andaluz de F\'isica de Part\'iculas Elementales,\\
           Universidad de Granada, 18071 Granada,Spain  }
}

\author{M\'onica Borunda}{
  address={Departamento de F\'isica Te\'orica y del Cosmos\\ 
           and Centro Andaluz de F\'isica de Part\'iculas Elementales,\\
           Universidad de Granada, 18071 Granada,Spain }
}

\author{Bert Janssen}{
  address={Departamento de F\'isica Te\'orica y del Cosmos\\
           and Centro Andaluz de F\'isica de Part\'iculas Elementales,\\
           Universidad de Granada, 18071 Granada,Spain }
  %,altaddress={<author1 address>} % additional visiting address
}

\begin{abstract}
We compare the metric and the Palatini formalism to obtain the Einstein equations in the presence 
of higher-order curvature corrections that consist of contractions of the Riemann tensor, but not 
of its derivatives. We find that in general the two formalisms are not equivalent and that the 
set of solutions of the Palatini equations is a non-trivial subset of the solutions of the metric 
equations. However we also argue that for Lovelock gravities, the equivalence of the two formalism 
holds completely and give an explanation of why it holds precisely for these theories.\footnote{
Talk given by M. Borunda; this article is a summary of \cite{BJB}.} 
\end{abstract}

\maketitle

%%%%%%%%%%%%%%%%%%%%%%%%%%%%%%%%%%%%%%%%%%%%
%% MAINMATTER
%%%%%%%%%%%%%%%%%%%%%%%%%%%%%%%%%%%%%%%%%%%%
%\section{Introduction}

One of the main lessons of General Relativity is that spacetime is a dynamical entity, with 
physical degrees of freedom, just like the matter and field content. Mathematically, spacetime is 
then described by a $D$-dimensional manifold, equipped with a metric $g_{\mu\nu}$ and a connection 
$\Gamma_{\mu\nu}^\rho$, whose dynamics is described by the Principle of Minimal Action. In 
differential geometry, the metric and the connection are two independent quantities and only 
assuming the connection to be symmetric ($\Gamma_{\mu\nu}^\rho = \Gamma_{\nu\mu}^\rho$) and metric 
compatible ($\nabla_\mu g_{\nu\rho}=0$), the connection is uniquely determined by the metric 
components, yielding the so-called Levi-Civita connection,
\begin{equation}
\Gamma_{\mu\nu}^\rho = \tfrac{1}{2} g^{\rho\lambda} \Bigl( \part_\mu g_{\lambda\nu} 
                                         + \part_\mu g_{\mu\lambda}
                                             - \part_\lambda g_\mn \Bigr).
\label{Levi-Civita}
\end{equation}
In General Relativity, one usually (tacitly) assumes, mostly due to simplicity and uniqueness 
arguments, that the Levi-Civita connection describes correctly the physics in Nature and hence 
that the metric is the only dynamical variable in the theory. However there also exists a 
mathematically more rigorous argument, called the Palatini formalism \cite{Palatini}, to prefer 
the Levi-Civita connection above more general ones, at least for Einstein gravity. 

The argument goes as follows: Consider the Einstein-Hilbert action as a functional of the metric 
and an arbitrary connection $\Gamma_\mn^\rho$, independent of the metric,
\be
S (g, \Gamma)= \int d^Dx \ \sqrtg \ g^\mn R_\mn(\Gamma),
\label{Einstein-Hilbert}
\ee 
where $R_\mn (\Gamma)$ is the Ricci tensor associated to $\Gamma_\mn^\rho$. The equation of motion
of the metric gives directly the Einstein equation for arbitrary connections, 
$R_\mn (\Gamma)- \half g_\mn R(\Gamma) = 0$, while the one for the connection identifies it
as being Levi-Civita. Not only this set of equation is equivalent to 
the Einstein equations obtained from metric formalism (i.e. supposing Levi-Civita and varying 
w.r.t the metric), but it also identifies the Levi-Civita connection as a minimum of the action, 
rather than a convenient choice.

Though the Einstein-Hilbert action is the natural choice for an action for gravity in four 
dimensions, there is no reason to exclude higher-curvature terms in higher dimensions. 
Higher-curvature corrections appear naturally in string theory, but will in general give rise to 
higher-order differential equations, thus introducing ghosts in the theory. However  Lanczos 
\cite{Lanczos}, and later Lovelock \cite{Lovelock}, presented a family of Lagrangians, called 
Lovelock gravities, that give rise to only second-order (and hence ghost free) Einstein equations.
For every order in the curvature terms there is a unique Lovelock Lagrangian, that therefore can 
be considered as the natural extensions of the Einstein-Hilbert action to higher dimensions.
A natural question to ask now is whether the Palatini formalism holds in the presence of 
general higher-curvature terms. 

Consider a general gravitational action that is a functional of the 
metric and (contractions of) the Riemann tensor, but not of its derivatives,
\be
S= \int d^D x \sqrtg \ \cL(g_\mn, R_\mnr{}^\lambda).
\label{action}
\ee 
In the metric formalism, the Einstein equation is obtained by varying this action $S(g)$
w.r.t. the explicit metric and the metrics inside the Riemann tensors, via the chain rule, 
\bea
\delta R_\mnr{}^\lambda \ = \ 2 \nabla_{[\mu} (\delta \Gamma_{\nu]\rho}^\lambda) 
                            %- \nabla_\nu (\delta \Gamma_{\mu\rho}^\lambda)
\hsp{1cm}
%, \nnw
\delta \Gamma_\mn^\rho \ = \ \half g^{\rho\lambda} \Bigl[
            2 \nabla_{(\mu} (\delta g_{\nu)\lambda}) %+  \nabla_\nu (\delta g_{\mu\lambda}) 
                          - \nabla_\lambda (\delta g_\mn) \Bigr],
\label{variation}
\eea  
such that the gravitational tensor $H_\mn \equiv \sqrtg^{-1} (\delta S(g)/\delta g^\mn)$ is then 
given by
\bea
H_\mn &=& \frac{\delta \cL}{\delta g^\mn} \ - \ \half g_\mn \cL  
  \ + \ \half [\nabla_\alpha, \nabla_\beta] 
      \Bigl(\frac{\delta \cL}{\delta R_{\alpha\beta\rho}{}^\lambda} \Bigr) 
                       g_{\rho(\mu} \delta_{\nu)}^\lambda 
\nnw
&-& \nabla_\rho \nabla_\alpha \Bigl(\frac{\delta \cL}{\delta R_{\alpha\beta\rho}{}^\lambda} \Bigr)
                 g_{\beta(\mu}\delta_{\nu)}^\lambda 
\ + \ \nabla^\lambda \nabla_\alpha \Bigl(\frac{\delta \cL}{\delta R_{\alpha\beta\rho}{}^\lambda} 
              \Bigr) g_{\beta(\mu} g_{\nu)\rho}.  
\label{H}
\eea
On the other hand, in the Palatini formalism the equations of motion of the metric and the 
connection generate a gravitational tensor $\cH_\mn$ and a connection tensor $\cK_\rho^\mn$ as
\be
\cH_\mn = \tfrac{1}{\sqrtg} \frac{\delta S(g, \Gamma)}{\delta g^\mn}, 
\hsp{1cm}
\cK^\mn_\rho = \tfrac{1}{\sqrtg} \frac{\delta S(g, \Gamma)}{\delta \Gamma_\mn^\rho}. 
\ee 

 In general the expressions for $\cH_\mn$ and $\cK_\rho^\mn$ are complicated, due to
the fact that for general connections the Riemann tensor has less symmetries than in the 
Levi-Civita case. However, since we are interested in comparing the Palatini formalism with the 
metric case, we will take the Levi-Civita connection as an Ansatz, substitute it in the 
$\cH_\mn$ and $\cK_\rho^\mn$ and compare the tensors with $H_\mn$. Imposing the Levi-Civita 
connection simplifies the expressions to
\bea
\cH_\mn \ = \ \frac{\delta \cL}{\delta g^\mn} \ - \ \half g_\mn \cL, \hsp{1cm}
\cK^{\mu\rho}_\lambda \ = \  \nabla_\nu \Bigl[ 
           \Bigl(\frac{\delta \cL}{\delta R_{\rho\nu\mu}{}^\lambda}\Bigr)
         - \Bigl(\frac{\delta \cL}{\delta R_{\nu\rho\mu}{}^\lambda}\Bigr)  \Bigr].
\label{cH,K}
\eea
The point now is to compare the equations obtained in the different formalisms. From (\ref{H})
and (\ref{cH,K}) it is clear that the main difference between $H_\mn$ and $\cH_\mn$ is the 
absence of second derivative terms in the latter. It is then natural to try to write the 
difference between these two tensors in terms of derivatives of $\cK_{\mn}^\lambda$. Indeed, we 
have that \cite{BJB}
\be
H_\mn = \cH_\mn - \half \nabla_\rho \cK^\rho_{(\mn)} 
               + \half g_{\lambda\mu} \nabla^\rho \cK_{(\nu\rho)}^\lambda
               + \half g_{\lambda\nu} \nabla^\rho \cK_{(\mu\rho)}^\lambda.
\label{H=H+K}
\ee 
This result has been derived earlier in Ref. \cite{CMQ}, but through a completely different 
approach, imposing the Levi-Civita connection via a Lagrange multiplier, such that the connection 
is not really an independent field, while we first derived the independent equations of motion and 
then substituted Levi-Civita as an Anstaz. It might seem remarkable at first sight that both 
methods yield the same results.

The question now arises whether the two formalisms are really equivalent, {\it i.e.}, whether 
any solution of one set also solves the equations of the other set. It will be clear that, in 
general, equation (\ref{H=H+K}) states that the Palatini formalism is contained within the metric 
formalism: any solution of the equations of motion in the Palatini formalism 
\bea
\cH_\mn = -\kappa T_\mn, \hsp{2cm}
\cK_{(\mu\nu)}^\lambda = 0,
\label{palatiniEOM}
\eea
is  also a solution of the Einstein equation in the metric formalism, 
%$H_\mn = -\kappa T_\$, 
which using (\ref{H=H+K}) can be written as
\be
\cH_\mn - \half \nabla_\rho \cK^\rho_{(\mn)} 
               + \half g_{\lambda(\mu} \nabla^\rho \cK_{\nu)\rho}^\lambda
               + \half g_{\lambda(\nu} \nabla^\rho \cK_{\mu)\rho}^\lambda = -\kappa T_\mn.
\label{metricEOM}
\ee
The opposite however is not necessarily true: in a general solution the different terms in the 
left-hand side of (\ref{metricEOM}) will conspire to satisfy the equation, rather than 
spontaneously decompose along the lines of (\ref{palatiniEOM}). In general the solutions of the 
Palatini formalism (supposing Levi-Civita) is a non-trivial subset of the solutions of the metric 
formalism. In \cite{BJB} specific examples are given of solutions of (\ref{metricEOM}) that do
not solve (\ref{palatiniEOM}).

A natural question then is to ask under which conditions solutions of the metric formalism also 
solve the Palatini equations and what the physical meaning of these conditions is. From 
(\ref{palatiniEOM}) we see that a necessary and sufficient condition is that the connection tensor 
vanishes, while from (\ref{cH,K}) we see that $\cK^{\mu\rho}_\lambda$ has the structure of a 
divergence, 
\be
\cK^{\mu\rho}_\lambda = \nabla_\nu B^{\nu\mu\rho}{}_\lambda.
\label{conserv}
\ee 
The vanishing of $\cK^{\mu\rho}_\lambda$ 
therefore implies a conserved current $B^{\nu\mu\rho}{}_\lambda$, which depends on the Lagrangian 
under consideration. Since $B^{\nu\mu\rho}{}_\lambda$ can be written in terms of contractions of the 
Riemann tensor, the connection equation imposes certain extra symmetry requirements on the metric 
and only those solutions of the metric equations that posses this symmetry are also solutions of 
the Palatini formalism. 

However it is also clear that for Lagrangians for which $\cK^\rho_{(\mn)}$ is identically zero, 
equations (\ref{palatiniEOM}) and (\ref{metricEOM}) happen to become equivalent, as the 
equations (\ref{metricEOM}) reduce to (\ref{palatiniEOM}). In fact a family of such Lagrangians 
exists and they turn out to be precisely the Lovelock gravities \cite{ESJ}. 
At first sight it might seems surprising that precisely the Lagrangians that yield only 
second-order differential equations (and hence are ghost-free), happen to be also the ones for 
which the Palatini formalism is completely equivalent to the metric formalism. However there is 
an easy way to see that these two properties of Lovelock Lagrangians are in fact closely related. 

Let us first have a closer look at the relation between the different gravitational tensors 
and the connection tensor, as put in equation (\ref{H=H+K}). Where by definition $H_\mn$ is the 
variation with respect to both the explicit metrics and metrics inside curvature tensors,
$\cH_\mn$ and $\cK_\mn^\rho$ are the variations of the action (\ref{action}) w.r.t. explicit 
metrics and connections respectively. Since the connections appear precisely in the 
curvature tensors, {\it i.e} in the places that in the metric formalism contain the implicit 
metrics, it is clear that the relation (\ref{H=H+K}) is therefore nothing more than a reflexion 
of the variation via the chain rule (\ref{variation}) in the metric formalism.

On the other hand, note that it is exactly the $\nabla \cK$ terms that give rise to higher-order 
derivatives in the Einstein equations, due to the non-linearity of the metric in the curvature 
tensors. Hence Lagrangians with identically vanishing connection tensors not only will have an
equivalence between the metric and the Palatini formalism, but will also yield only second-order
differential equations. But of course these Lagrangians are uniquely identified as Lovelock 
gravities, such that the second-order differential equations and equivalence of the Palatini and 
metric formalisms are in fact two aspects of the same property.

\vsp{1cm}
%%%%%%%%%%%%%%%%%%%%%%%%%%%%%%%%%%%%%%%%%%%%%%%%
%% BACKMATTER
%%%%%%%%%%%%%%%%%%%%%%%%%%%%%%%%%%%%%%%%%%%%%%%%

\begin{theacknowledgments}
We wish to thank Q. Exirifard and M.M. Sheikh-Jabbari for enlightening discussions. 
The work of M.B.G.~and B.J.~is done as part of the program {\sl Ram\'on y Cajal} and the work 
of M.B. is done as part of the program {\sl Juan de la Cierva}, both of the Ministerio de 
Educaci\'on y Ciencias (M.E.C.) of Spain. The authors are also partially supported by the M.E.C. 
under contract FIS 2007-63364 and by the Junta de Andaluc\'{\i}a group FQM 101.

\end{theacknowledgments}

%%%%%%%%%%%%%%%%%%%%%%%%

\bibliographystyle{aipproc}   % if natbib is available

\begin{thebibliography}{9}

\bibitem{BJB} M. Borunda, B. Janssen and M. Bastero-Gil, JCAP  0811 (2008) 008.

\bibitem{Palatini} A. Palatini, Rend. Circ. Mat. Palermo 43, (1919) 203.

\bibitem{Lanczos} C. Lanczos, Ann. Math. 39 (1938) 842. 
         
\bibitem{Lovelock} D. Lovelock, Aequationes Math. 4 (1970) 127; 
                   D. Lovelock, J. Math. Phys. 12 (1971) 498.

%\bibitem{Z-Z} B. Zwiebach, Phys. Lett B 156 (1985) 315; B. Zumino, Phys. Rep.  137 (1986) 109.

\bibitem{CMQ} S. Cotsakis, J. Miritzis and L. Querella, J. Math. Phys. 40 (1999) 3063.

\bibitem{ESJ} Q. Exirifard and M. M. Sheikh-Jabbari, Phys. Lett. B661 (2008) 158.  


\end{thebibliography}
%\bibliographystyle{aipprocl} % if natbib is missing

%%%%%%%%%%%%%%%%%%%%%%%%%%%%
\end{document}